\documentstyle[12pt]{article}

\expandafter\ifx\csname amssym.def\endcsname\relax \else\endinput\fi
\expandafter\edef\csname amssym.def\endcsname{%
       \catcode`\noexpand\@=\the\catcode`\@\space}
\catcode`\@=11
\def\undefine#1{\let#1\undefined}
\def\newsymbol#1#2#3#4#5{\let\next@\relax
 \ifnum#2=\@ne\let\next@\msafam@\else
 \ifnum#2=\tw@\let\next@\msbfam@\fi\fi
 \mathchardef#1="#3\next@#4#5}
\def\mathhexbox@#1#2#3{\relax
 \ifmmode\mathpalette{}{\m@th\mathchar"#1#2#3}%
 \else\leavevmode\hbox{$\m@th\mathchar"#1#2#3$}\fi}
\def\hexnumber@#1{\ifcase#1 0\or 1\or 2\or 3\or 4\or 5\or 6\or 7\or 8\or
 9\or A\or B\or C\or D\or E\or F\fi}

\ifcase\@ptsize
  \font\tenmsa=msam10
  \font\sevenmsa=msam7
  \font\fivemsa=msam5
\or
  \font\tenmsa=msam10  scaled \magstephalf
  \font\sevenmsa=msam7 scaled \magstephalf
  \font\fivemsa=msam5  scaled \magstephalf
\or
  \font\tenmsa=msam10  scaled \magstep1
  \font\sevenmsa=msam7 scaled \magstep1
  \font\fivemsa=msam5  scaled \magstep1
\fi

\newfam\msafam
\textfont\msafam=\tenmsa
\scriptfont\msafam=\sevenmsa
\scriptscriptfont\msafam=\fivemsa
\edef\msafam@{\hexnumber@\msafam}
\mathchardef\dabar@"0\msafam@39
\def\dashrightarrow{\mathrel{\dabar@\dabar@\mathchar"0\msafam@4B}}
\def\dashleftarrow{\mathrel{\mathchar"0\msafam@4C\dabar@\dabar@}}

\def\ulcorner{\delimiter"4\msafam@70\msafam@70 }
\def\urcorner{\delimiter"5\msafam@71\msafam@71 }
\def\llcorner{\delimiter"4\msafam@78\msafam@78 }
\def\lrcorner{\delimiter"5\msafam@79\msafam@79 }
\def\yen{{\mathhexbox@\msafam@55 }}
\def\checkmark{{\mathhexbox@\msafam@58 }}
\def\circledR{{\mathhexbox@\msafam@72 }}
\def\maltese{{\mathhexbox@\msafam@7A }}

\ifcase\@ptsize
  \font\tenmsb=msbm10
  \font\sevenmsb=msbm7
  \font\fivemsb=msbm5
\or
  \font\tenmsb=msbm10  scaled \magstephalf
  \font\sevenmsb=msbm7 scaled \magstephalf
  \font\fivemsb=msbm5  scaled \magstephalf
\or
  \font\tenmsb=msbm10  scaled \magstep1
  \font\sevenmsb=msbm7 scaled \magstep1
  \font\fivemsb=msbm5  scaled \magstep1
\fi

\newfam\msbfam
\textfont\msbfam=\tenmsb
\scriptfont\msbfam=\sevenmsb
\scriptscriptfont\msbfam=\fivemsb
\edef\msbfam@{\hexnumber@\msbfam}
\def\Bbb#1{{\fam\msbfam\relax#1}}
\def\widehat#1{\setbox\z@\hbox{$\m@th#1$}%
 \ifdim\wd\z@>\tw@ em\mathaccent"0\msbfam@5B{#1}%
 \else\mathaccent"0362{#1}\fi}
\def\widetilde#1{\setbox\z@\hbox{$\m@th#1$}%
 \ifdim\wd\z@>\tw@ em\mathaccent"0\msbfam@5D{#1}%
 \else\mathaccent"0365{#1}\fi}

\ifcase\@ptsize
  \font\teneufm=eufm10
  \font\seveneufm=eufm7
  \font\fiveeufm=eufm5
\or
  \font\teneufm=eufm10   scaled \magstephalf
  \font\seveneufm=eufm7  scaled \magstephalf
  \font\fiveeufm=eufm5   scaled \magstephalf
\or
  \font\teneufm=eufm10   scaled \magstep1
  \font\seveneufm=eufm7  scaled \magstep1
  \font\fiveeufm=eufm5   scaled \magstep1
\fi

\newfam\eufmfam
\textfont\eufmfam=\teneufm
\scriptfont\eufmfam=\seveneufm
\scriptscriptfont\eufmfam=\fiveeufm
\def\frak#1{{\fam\eufmfam\relax#1}}

\csname amssym.def\endcsname

\expandafter\ifx\csname pre amssym.tex at\endcsname\relax \else \endinput\fi
\expandafter\chardef\csname pre amssym.tex at\endcsname=\the\catcode`\@
\catcode`\@=11
\newsymbol\boxdot 1200
\newsymbol\boxplus 1201
\newsymbol\boxtimes 1202
\newsymbol\square 1003
\newsymbol\blacksquare 1004
\newsymbol\centerdot 1205
\newsymbol\lozenge 1006
\newsymbol\blacklozenge 1007
\newsymbol\circlearrowright 1308
\newsymbol\circlearrowleft 1309
\undefine\rightleftharpoons
\newsymbol\rightleftharpoons 130A
\newsymbol\leftrightharpoons 130B
\newsymbol\boxminus 120C
\newsymbol\Vdash 130D
\newsymbol\Vvdash 130E
\newsymbol\vDash 130F
\newsymbol\twoheadrightarrow 1310
\newsymbol\twoheadleftarrow 1311
\newsymbol\leftleftarrows 1312
\newsymbol\rightrightarrows 1313
\newsymbol\upuparrows 1314
\newsymbol\downdownarrows 1315
\newsymbol\upharpoonright 1316
 
\newsymbol\downharpoonright 1317
\newsymbol\upharpoonleft 1318
\newsymbol\downharpoonleft 1319
\newsymbol\rightarrowtail 131A
\newsymbol\leftarrowtail 131B
\newsymbol\leftrightarrows 131C
\newsymbol\rightleftarrows 131D
\newsymbol\Lsh 131E
\newsymbol\Rsh 131F
\newsymbol\rightsquigarrow 1320
\newsymbol\leftrightsquigarrow 1321
\newsymbol\looparrowleft 1322
\newsymbol\looparrowright 1323
\newsymbol\circeq 1324
\newsymbol\succsim 1325
\newsymbol\gtrsim 1326
\newsymbol\gtrapprox 1327
\newsymbol\multimap 1328
\newsymbol\therefore 1329
\newsymbol\because 132A
\newsymbol\doteqdot 132B
 
\newsymbol\triangleq 132C
\newsymbol\precsim 132D
\newsymbol\lesssim 132E
\newsymbol\lessapprox 132F
\newsymbol\eqslantless 1330
\newsymbol\eqslantgtr 1331
\newsymbol\curlyeqprec 1332
\newsymbol\curlyeqsucc 1333
\newsymbol\preccurlyeq 1334
\newsymbol\leqq 1335
\newsymbol\leqslant 1336
\newsymbol\lessgtr 1337
\newsymbol\backprime 1038
\newsymbol\risingdotseq 133A
\newsymbol\fallingdotseq 133B
\newsymbol\succcurlyeq 133C
\newsymbol\geqq 133D
\newsymbol\geqslant 133E
\newsymbol\gtrless 133F
\newsymbol\sqsubset 1340
\newsymbol\sqsupset 1341
\newsymbol\vartriangleright 1342
\newsymbol\vartriangleleft 1343
\newsymbol\trianglerighteq 1344
\newsymbol\trianglelefteq 1345
\newsymbol\bigstar 1046
\newsymbol\between 1347
\newsymbol\blacktriangledown 1048
\newsymbol\blacktriangleright 1349
\newsymbol\blacktriangleleft 134A
\newsymbol\vartriangle 134D
\newsymbol\blacktriangle 104E
\newsymbol\triangledown 104F
\newsymbol\eqcirc 1350
\newsymbol\lesseqgtr 1351
\newsymbol\gtreqless 1352
\newsymbol\lesseqqgtr 1353
\newsymbol\gtreqqless 1354
\newsymbol\Rrightarrow 1356
\newsymbol\Lleftarrow 1357
\newsymbol\veebar 1259
\newsymbol\barwedge 125A
\newsymbol\doublebarwedge 125B
\undefine\angle
\newsymbol\angle 105C
\newsymbol\measuredangle 105D
\newsymbol\sphericalangle 105E
\newsymbol\varpropto 135F
\newsymbol\smallsmile 1360
\newsymbol\smallfrown 1361
\newsymbol\Subset 1362
\newsymbol\Supset 1363
\newsymbol\Cup 1264
 
\newsymbol\Cap 1265
 
\newsymbol\curlywedge 1266
\newsymbol\curlyvee 1267
\newsymbol\leftthreetimes 1268
\newsymbol\rightthreetimes 1269
\newsymbol\subseteqq 136A
\newsymbol\supseteqq 136B
\newsymbol\bumpeq 136C
\newsymbol\Bumpeq 136D
\newsymbol\lll 136E
 
\newsymbol\ggg 136F
 
\newsymbol\circledS 1073
\newsymbol\pitchfork 1374
\newsymbol\dotplus 1275
\newsymbol\backsim 1376
\newsymbol\backsimeq 1377
\newsymbol\complement 107B
\newsymbol\intercal 127C
\newsymbol\circledcirc 127D
\newsymbol\circledast 127E
\newsymbol\circleddash 127F
\newsymbol\lvertneqq 2300
\newsymbol\gvertneqq 2301
\newsymbol\nleq 2302
\newsymbol\ngeq 2303
\newsymbol\nless 2304
\newsymbol\ngtr 2305
\newsymbol\nprec 2306
\newsymbol\nsucc 2307
\newsymbol\lneqq 2308
\newsymbol\gneqq 2309
\newsymbol\nleqslant 230A
\newsymbol\ngeqslant 230B
\newsymbol\lneq 230C
\newsymbol\gneq 230D
\newsymbol\npreceq 230E
\newsymbol\nsucceq 230F
\newsymbol\precnsim 2310
\newsymbol\succnsim 2311
\newsymbol\lnsim 2312
\newsymbol\gnsim 2313
\newsymbol\nleqq 2314
\newsymbol\ngeqq 2315
\newsymbol\precneqq 2316
\newsymbol\succneqq 2317
\newsymbol\precnapprox 2318
\newsymbol\succnapprox 2319
\newsymbol\lnapprox 231A
\newsymbol\gnapprox 231B
\newsymbol\nsim 231C
\newsymbol\ncong 231D
\newsymbol\diagup 231E
\newsymbol\diagdown 231F
\newsymbol\varsubsetneq 2320
\newsymbol\varsupsetneq 2321
\newsymbol\nsubseteqq 2322
\newsymbol\nsupseteqq 2323
\newsymbol\subsetneqq 2324
\newsymbol\supsetneqq 2325
\newsymbol\varsubsetneqq 2326
\newsymbol\varsupsetneqq 2327
\newsymbol\subsetneq 2328
\newsymbol\supsetneq 2329
\newsymbol\nsubseteq 232A
\newsymbol\nsupseteq 232B
\newsymbol\nparallel 232C
\newsymbol\nmid 232D
\newsymbol\nshortmid 232E
\newsymbol\nshortparallel 232F
\newsymbol\nvdash 2330
\newsymbol\nVdash 2331
\newsymbol\nvDash 2332
\newsymbol\nVDash 2333
\newsymbol\ntrianglerighteq 2334
\newsymbol\ntrianglelefteq 2335
\newsymbol\ntriangleleft 2336
\newsymbol\ntriangleright 2337
\newsymbol\nleftarrow 2338
\newsymbol\nrightarrow 2339
\newsymbol\nLeftarrow 233A
\newsymbol\nRightarrow 233B
\newsymbol\nLeftrightarrow 233C
\newsymbol\nleftrightarrow 233D
\newsymbol\divideontimes 223E
\newsymbol\varnothing 203F
\newsymbol\nexists 2040
\newsymbol\Finv 2060
\newsymbol\Game 2061
\newsymbol\mho 2066
\newsymbol\eth 2067
\newsymbol\eqsim 2368
\newsymbol\beth 2069
\newsymbol\gimel 206A
\newsymbol\daleth 206B
\newsymbol\lessdot 236C
\newsymbol\gtrdot 236D
\newsymbol\ltimes 226E
\newsymbol\rtimes 226F
\newsymbol\shortmid 2370
\newsymbol\shortparallel 2371
\newsymbol\smallsetminus 2272
\newsymbol\thicksim 2373
\newsymbol\thickapprox 2374
\newsymbol\approxeq 2375
\newsymbol\succapprox 2376
\newsymbol\precapprox 2377
\newsymbol\curvearrowleft 2378
\newsymbol\curvearrowright 2379
\newsymbol\digamma 207A
\newsymbol\varkappa 207B
\newsymbol\Bbbk 207C
\newsymbol\hslash 207D
\undefine\hbar
\newsymbol\hbar 207E
\newsymbol\backepsilon 237F
\catcode`\@=\csname pre amssym.tex at\endcsname

\def\Box{\hbox{\vrule height1ex\kern-0.4pt
\vbox to 1ex{\hrule width1ex\vfil\hrule width1ex}\kern-0.4pt\vrule height1ex}}
\newcommand{\sqr}[2]{{{\vcenter{\vbox{\hrule height.#2pt
\hbox{\vrule width.#2pt height#1pt \kern#1pt
\vrule width.#2pt}
\hrule height.#2pt}}}}}

\newcommand{\be}{\begin{equation}}

\newcommand{\ee}{\end{equation}}

\newcommand{\dl}{\delta}
\newcommand{\ep}{\epsilon}

\newcommand{\lm}{\lambda}

\newcommand{\rh}{\rho}
\newcommand{\sg}{\sigma}

\newcommand{\phv}{\varphi}

\newcommand{\ps}{\psi}

\newcommand{\om}{\omega}

\newcommand{\raw}{\rightarrow}
\newcommand{\A}{\frak A}
\newcommand{\B}{\frak B}

\renewcommand{\H}{\mbox{$\cal H$}}

 \renewcommand{\ll}{\label}
 \newcommand{\fn}{\footnote}
 \renewcommand{\sp}{\samepage}

\newcommand{\notp}{p \kern-.48em /}

\newcommand{\bea}{\begin{eqnarray}}
\newcommand{\eea}{\end{eqnarray}}
\newcommand{\ot}{\otimes}

\topmargin = - 0.5 cm
\newcommand{\up}{\uparrow}
\newcommand{\dow}{\downarrow}
\newcommand{\Up}{\uparrow\ldots\uparrow}
\newcommand{\Do}{\downarrow\ldots\downarrow}
\newcommand{\Z}{{\cal Z}}
\begin{document}
\setlength{\baselineskip}{1.5\baselineskip}
\thispagestyle{empty}
\title{Observation and superselection in quantum mechanics\thanks{To appear in
{\em Studies in
History and Philosophy of Modern Physics} (1995)}} \author{
N.P.~Landsman\thanks{Alexander von
Humboldt Fellow and  S.E.R.C. Advanced Research Fellow; permanent address:
Department of Applied
Mathematics and Theoretical Physics, University of Cambridge, Silver Street,
Cambridge CB3 9EW, United Kingdom}\\ \mbox{}\hfill \\ II. Institut f\"{u}r
Theoretische Physik,
Universit\"{a}t Hamburg\\ Luruper Chaussee 149, 22761 Hamburg, Germany }
\maketitle
\begin{abstract}
We attempt to clarify the main conceptual issues in approaches to
`objectification' or `measurement'
in quantum mechanics which are based on superselection rules.
Such approaches  venture to derive the emergence of classical `reality'
relative to a class of observers; those believing that the classical world
exists intrinsically and
absolutely are advised against reading this paper.

The prototype approach (Hepp) where superselection sectors are assumed in the
state space of the
apparatus is shown to be untenable. Instead, one should couple system and
apparatus to an environment,
and postulate superselection rules for the latter. These are motivated by the
locality of any
observer or other (actual or virtual) monitoring system. In this way
`environmental' solutions to
the measurement problem (Zeh, Zurek) become consistent and acceptable, too.
Points of contact with
the modal interpretation are briefly discussed.

We propose a minimal value attribution to observables  in theories with
superselection rules,
 in which only central   observables have properties.
In particular, the eigenvector-eigenvalue link is dropped. This is mainly
motivated by Ockham's
razor.
 \end{abstract}
 \newpage
\section{Introduction}
The original title of this paper was ``To observe is to not observe'', but it
was pointed out to the
author that this represented a contradiction.  Our first aim  is to discuss
certain lines
of criticism that have been, or could be, leveled against  resolutions of the
measurement problem in
quantum mechanics which essentially rely on the (algebraic) theory of
superselection rules. Secondly,
we will indicate how this theory may be combined with more recent ideas on
decoherence and
apparatus-environment coupling in order to    counter the more  pertinent
critique.  Thus we will
arrive at the following  point of view: the essence of a   `measurement',
`fact', or `event' in
quantum mechanics lies in the   non-observation, or irrelevance, of a certain
part of the system in
question. The latter may well be the universe as a whole; one is not forced to
make a `Heisenberg
cut' between system and observer, and in our analysis the observer (or IGUS =
Information Gathering
and Utilizing System in modern parlance  (Gell-Mann and Hartle, 1990, 1993);
whenever we speak of an
observer in what follows, the reader may add `or IGUS') relative to which the
notion of
non-observation or irrelevance is defined may be regarded as part of the
system, and may be described
by quantum mechanics if necessary. Without such irrelevance of some part of the
system the notion of
a fact (etc.) is meaningless in quantum mechanics, or, put differently, there
can only be events once
a specific algebra of `observables' has been singled out. Any event that
`happens' only comes into
existence relative to such a choice of `observables', and on the assumption of
the ignorance
interpretation of mixed states.   A world without parts declared or forced to
be irrelevant is a
world without facts - such a world may be preferable to ours. As we shall see,
in practice facts owe
their existence to the locality of the observer (a point of view the author
learned from H.D. Zeh).

The measurement problem  (cf.\   Busch et al.\ (1991), van Fraassen (1991) for
an extensive
discussion) is a special case of the enigma of classical behaviour within
quantum mechanics. The precise formulation of the problem depends on the
formalism one uses, and on
the interpretative rules connecting the formalism to the world. The fine kettle
of fish is most
evident if in the usual (von Neumann) liturgy one assumes a one-to-one
correspondence between the
physical properties of a system (in the sense of value attributions to
observables) and its states.
For in that case the formalism predicts the existence of states which seem to
never occur.
Such states are superpositions of eigenstates of  operators which are
`classical' in the sense that
the corresponding observables are empirically found to  always possess sharp
values.

The aim of this Introduction is to specify where the theory of superselection
rules (on our reading)
stands within the debate on the foundations of quantum mechanics, and to
introduce the essential
points of this theory with its physical interpretation. In section 2 we study a
model, and we
identify its main diseases in section 3. The best cure
is
investigated in section 4 in the form of the introduction of the environment
into the problem.
Section 5 then discusses points of contact with the modal interpretation of
quantum mechanics.
The final section contains some self-criticism, as well as a summary of our
approach in the form of
a question- and answer session.

In order to explain what we accept as chivalrous criticism of superselection
approaches to the
measurement problem, and which type of argument we reject as mock critique, we
recall that in the
context of quantum mechanics there exist two radically different views on the
nature of `classical'
reality (for a good discussion cf.\ D'Espagnat (1990), Tsirelson (1994)). The
difference between the
two has been expressed in Khalfin and Tsirelson (1992, p.\ 904) by saying that
``the
`optimists' investigate the emergence of classical reality relative to a class
of observers, whereas
the `pessimists' acknowledge only absolute (independent) classical reality''.
To elaborate on this
point, we  distinguish between two further positions, which have both been
advocated as `realism'.

 The first position, which we call {\em A-realism},
maintains that there exists a real world independently of the observer, and
that one can make
objective, observer-independent statements about it. This creed is meant to be
contrasted with
idealism, solipsism, and the like. It is very broad, and further subdivisions
arise once an
A-realist specifies how (s)he relates the alleged real world to our
observations, and what the aim
of science should be. Thus almost opposite extremes such as naive realism and
constructive empiricism
(which is presented as an anti-realist pursuasion  in van Fraassen (1980)) both
fall under
A-realism.

 This position has been attacked by some of the pioneers of quantum
mechanics, and has been claimed to be inconsistent with it;  the existence of
the modal interpretation
of quantum mechanics  (van Fraassen, 1991, Kochen, 1985, Healey, 1989, Dieks,
1994a,b) shows that this
 latter claim is false. We see nothing objectionable to A-realism, and  will
adopt it in this paper.

The second position, {\em B-realism}, is in fact a specialization of A-realism,
but in a good many
papers on the foundations of quantum mechanics it is confused with (usually
unqualified) `realism'
itself. It claims that
this postulated real and independently existing world coincides with, or at
least incorporates the
classical world of `events' and `facts' that we observe around us. One may
compare this with the
pre-Copernican world view: the earth appears to be at rest, the sun revolving
around it, and since it
looks so to us, it must be real and true, independently of us. Some Anglo-Saxon
more sympathetic to
B-realism might instead describe it as the application of G.E. Moore's
common-sense realism to the
interpretation of quantum mechanics. However, once it is realized that we
should try to understand why
we see things as we do, rather than  explaining why these things  (supposedly)
`are' the way we see
them, one cannot help feeling awkward with B-realism.

Whether or not an author advocating superselection- or decoherence-type
solutions to the measurement
problem is an A-realist, (s)he will definitely reject B-realism at least when
analyzing such
solutions, for it is the whole point of these approaches to show that under
certain conditions the
classical world emerges relative to, say, local observables. Hence (s)he is an
example of an optimist
in the sense of the above quote. From the point of view of a B-realist, such
solutions are at best
valid `For All Practical Purposes (FAPP)', and thus a large body of  criticism
on superselection- or
decoherence-approaches can be summarized simply by saying that  these
approaches do not  conform to
B-realism. We believe that this type of critique is based on a hallucination,
for
it is blind to the fact that the
notion of classical reality itself is only valid FAPP (and not quantum
mechanics, or resolutions of the measurement problem based on it).

In further motivating the  point of view opposite to B-realism, we
start by introducing some terminology.
Quantum mechanics may be fruitfully thought of as having a {\em kernel} as well
as a {\em user
interface}. The kernel relates to observer-independent aspects of the real
world. It primarily
consists of the mathematical formalism, involving either operators, states,
and transformations like time-evolution, or path integrals and an action
functional, or some other
mathematical machinery. Thus at the level of the kernel one may speak of
eigenvectors, eigenvalues,
and of mathematical expressions that are usually interpreted as   expectation
values of observables
or  transition probabilities.   This particular physical interpretation is very
delicate;
it evidently does not belong to the mathematical formalism, but
  in our opinion it is not even  part of the kernel at all (see below).
 In particular, it would be a mistake to assume the
so-called `eigenvector-eigenvalue link' (stating that an observable possesses a
value   if the
state vector of the system is an eigenstate of the corresponding operator) as
part of the kernel, and
neither do the Born probabilities make physical sense at this stage.

Secondarily, an A-realist will want to relate the mathematical formalism to the
observer-independent
world through interpretative rules which are part of the kernel. Such a
relation is  not
indispensable in order to confront the theory with observations, for one could
introduce the physical
interpretation of the mathematical formalism  at the level of the observer.
Indeed, the latter procedure is followed in the Copenhagen interpretation of
quantum mechanics, and
also in our exegesis of the theory of superselection rules given later on.
 But one would  clearly feel more comfortable if the empirical
content of the theory would follow from  a direct physical interpretation in
the kernel, amended by an
objective description of the observer. In our opinion, this ideal situation has
not (yet) been
achieved in quantum mechanics (the modal interpretation being an attempt in
that direction, cf.\
section 5 below).

The user interface specifies the connection between the real world and the
observer, regarded
as part of the world. Hence  (s)he/it is subject to the same laws (such as
quantum mechanics)
that govern the world. This interface is an objective account of those
ingredients of the world
which emerge only relative to the specification of an observer (or class of
observers).
Tautologically, observations in general are one example of such an ingredient.
More daringly,
we maintain that the entire classical world is another case in point.

The interpretation of mathematical expectation values in terms of measurements
(particularly the
eigenvector-eigenvalue rule), cross-sections, as well as more recent notions
such as the
probabilities of histories (Omn\`{e}s, 1992, Gell-Mann and Hartle, 1990, 1993),
properly belong to the
user interface. Evidently,  this interface cannot be created without a
specification of the user.
This is an elementary though crucial point: if we wish to explain from quantum
mechanics why the
world {\em appears to us} as it does, i.e., largely  classically, we should
expect this explanation
to come from the user interface, and therefore be contingent on what or who the
user is. (The
question why the world {\em is} largely classical seems to us to be as little
motivated as the
question why the present King of France is bald.)

A somewhat related  point was made   by Zurek  (1993, pp.\ 287-8): ``Thus, the
only sensible
subject of considerations aimed at the interpretation of quantum theory - that
is, at establishing
correspondence between the quantum formalism and the events perceived by us -
is the relation
between the universal state vector and the states of memory (records) of
somewhat special systems -
such as observers - which are, of necessity, perceiving the Universe from
within. It is the
inability to appreciate the consequences of this rather simple but fundamental
observation that has
led to such desperate measures as the search for an alternative to quantum
physics.''\fn{We do not
agree that such a search is a desperate move; the point is that it is misguided
when motivated
by the measurement problem.}
 Moreover,
the necessity of identifying the user interface of a theory as the source of
concepts naively
thought to belong to the kernel even applies to classical physics, for instance
in the problem of the
emergence of time in the context of  generally covariant field theories (cf.\
Barbour, 1994).

The theory of superselection rules is an important tool in attempts to make
these ideas precise. In
its modern algebraic version, this theory was created by Haag and Kastler
(1964), and the first
application to the measurement problem was performed by Hepp (1972), who
acknowledges that the
main ideas are due to Fierz and Jost; also cf.\  Bohm (1951), Gottfried (1966),
Jauch (1968) for
pioneering insights in this direction;
later work   is e.g.\ Wan (1980), Beltrametti and Cassinelli (1981). A recent
review, emphasizing
technical points and explaining relevant aspects of the mathematical apparatus
of operator algebras,
is Landsman (1991) (to which the present paper is complementary). In what
follows, we assume that the
reader is somewhat familiar with the basic ideas of this approach, but we will
keep technicalities to
a minimum.  Even if this familiarity is marginal or rusty, it will be easy to
understand the key
ideas on the basis of the example discussed in the next section.
The interpretation of the formalism we offer is quite different from the one
found in the literature
(where one is usually satisfied with straightforward operationalistic ideas). A
different motivation
from the one given here to justify superselection rules is in Breuer et al.\
(1994).

In fact, the main idea is very simple. In conventional quantum mechanics any
self-adjoint
operator on a given Hilbert space $\H$ is deemed an observable, and therefore
any unit vector in $\H$
corresponds to a pure state.
However, a realistic observer will not  actually monitor all conceivable
correlations in
the universe. Hence this setting may be modified by leaving out a certain set
of operators
from the `algebra of  beables' $\B(\H)$ (i.e., the algebra of all bounded
operators on $\H$), to
arrive at
 a remaining `algebra of
observables' $\A$ (which we assume to be a von Neumann algebra\fn{This means
that $\A$ consists of
bounded operators, contains the unit operator, and is closed in the weak
operator topology.
Hence it is closed in the uniform operator topology as well, so that each von
Neumann algebra is a
$C^*$-algebra if it is equipped with the latter topology.} for mathematical
convenience).
 In our interpretation, {\em the
truncation of the original set $\B(\H)$ of beables to a (much) smaller set of
observables $\A$
is
made by  the `user', who normally has little choice in doing so.} This
truncation therefore belongs to
the user interface of quantum mechanics.
 We then regard  a unit vector $\Psi$  in $\H$ as a state\fn{This is a linear
functional on $\A$
which is positive (i.e., $\ps(A^*A)\geq 0$) and normalized ($\ps({\Bbb I})=1$,
where $\Bbb I$ is the
unit operator in $\A$).}
 $\ps$ on $\A$.  Thus $\Psi$ determines a rule   $\psi$  telling each operator
$A$ in $\A$ what its expectation value $\psi(A)$ is, namely  $(A\Psi,\Psi)$.
But now such a state may
well be mixed\fn{A state $\ps$ is mixed if it may be decomposed as
$\ps=\lm\ps_1+(1-\lm)\ps_2$ with
$0<\lm<1$ and $\psi_1\neq \psi_2$.} on $\A$!

Therefore, we have to investigate the possible decompositions of $\Psi$ (or
$\ps$).
We make the simplifying assumption that the commutant\fn{This is the set of all
bounded operators on $\H$ which commute with all elements of $\A$.}
 $\A'$ of $\A$ on $\H$ is abelian (that is,
commutative).
This assumption\fn{Which implies that $\A$ must be a type I von Neumann
algebra.} amounts to the
statement that $\A$ is represented without multiplicities, i.e., redundant
repetitions of
information\fn{See footnote \ref{gen} for the general case.}.
 We say that $\Psi_1,\Psi_2\in\H$ lie in different superselection sectors (or,
briefly,
sectors) if $(A\Psi_1,\Psi_2)=0$ for all $A\in\A$\fn{Together with our
assumption, this implies
that the corresponding states $\ps_1,\ps_2$ define inequivalent representations
of $\A$.}.

In general, a mixed state $\ps$ on $\A$ may not have
 a unique extremal  decomposition\fn{An extremal  decomposition of a mixed
state is by definition a decomposition
into pure states. See Bratteli and Robinson (1987) for an exhaustive
mathematical account of the
decomposition theory of states on $C^*$-algebras, and cf.\ sect.\ 4.5 of
Landsman (1991) for the
general
 conditions under which the extremal decomposition is unique.}   $\psi=\sum_i
p_i \psi_i$ into pure
states $\psi_i$ (with the coefficients $p_i$ adding up to one)\fn{For
simplicity we suppress the
possibility of a direct integral decomposition of the state; the argument is
analogous in that
case.}.
  However,  we are in the special situation that the state $\ps$ on $\A$ is the
restriction to $\A\subset
\B(\H)$ of a  state defined by a vector $\Psi$ in $\H$.
In that case we can write $\Psi=\sum_i c_i\Psi_i$, with all the $\Psi_i$ lying
in different sectors,
 and this decomposition is unique. Moreover, our assumption that $\A'$ be
abelian implies that each
$\Psi_i$ corresponds to a pure state $\ps_i$ on $\A$.
 Hence with $p_i=|c_i|^2$ we obtain  the unique extremal decomposition
$\psi=\sum_i p_i \psi_i$.
 As will become clear shortly,  {\em
these coefficients $p_i$ are the Born probabilities}. The uniqueness of the
central decomposition is
crucial for this interpretation.

Indeed, at this point the formalism ought to be connected to observation. The
simplest way to do
so is to only assign properties to classical observables. Here a
  self-adjoint operator in $\A$ is called a classical observable if  it belongs
to the
centre $\Z(\A)$ of the algebra of observables $\A$ - this means that it
commutes with all other
elements of $\A$. Equivalently,
   eigenstates with different eigenvalues cannot be coherently superposed. Only
  classical
observables possess values, and all classical observables simultaneously assume
(sharp) values. In a
pure state $\ps$, the value of $A\in  \Z(\A)$ is simply $\ps(A)$.
To deal with mixed states (crucial in the measurement problem) we adopt the
ignorance interpretation
of mixed states which have a unique extremal decomposition; as shown above, if
the total universe is
in a pure state this uniqueness assumption can be justified\fn{If the universe
is in a mixed state
(whatever that may mean) we still recover the Born probabilities, as shown in
footnote \ref{gen}
below.}. Thus in a state
$\psi=\sum_i p_i \psi_i$ the classical observable $A$ has value $\ps_i(A)$ with
probability $p_i$.

 We have dropped the eigenvector-eigenvalue
link for general operators: for us, an eigenvector $\Psi_{a}$ of a given
operator $A\in\A$ corresponds
to a  possessed  (eigen) value $a$ of $A$ if and only if $A$ is classical.
Similarly, the Born
probability  $|(\Psi,\Psi_{a})|^2$  only has the usual meaning   if $A$ is
classical. In this circumstance,
each eigenvalue will be highly if not infinitely degenerate;   the probability
that the
observable $A$ has the value $a$ in the state $\Psi$ is the sum over all
pertinent Born
probabilities.
Our main reason for dropping the eigenvector-eigenvalue
link
 is Ockham's razor: as is well known (van Fraassen, 1991,
Healey, 1989) this link is empirically superfluous, and in our opinion it
results from misguided
attempts to assign properties to observables in a specific way, thus
introducing
the Born probabilities at the level of the
kernel of quantum mechanics. Further arguments in favour of our minimal value
attribution will be
presented in the course of this paper\fn{{\sp The first such argument comes
from states on $\A$ which
do not admit  a unique extremal decomposition. This may happen if the state on
$\B(\H)$ (whose
restriction to $\A$ is $\ps$) is itself mixed; alternatively, even a pure state
on $\B(\H)$
may restrict to such a  state on $\A$ if we drop the assumption that the
commutant $\A'$ on $\H$ is
abelian. (Lifting this assumption is necessary to describe certain states of
infinite systems, whose
algebras of observables always admit  representations which are not of type I.)

Any state $\ps$ has a unique
central decomposition $\ps=\sum_i p_i\om_i$ into `macroscopically pure' states.
 A macroscopically
pure state $\om$ on $\A$, also called a primary or a factorial state, may be
defined by the property
that any {\em discrete} decomposition  $\om=\sum_n \lm_n\om_n$ consists of
states $\om_n$ which all
lie in the same sector, up to multiplicity (that is, the GNS representations of
$\A$ defined by these
states are all unitarily equivalent up to multiplicity). If $\A$ is of type I
then $\om$ has discrete
{\em extremal} decompositions. (If $\A$ is not of type I the extremal
decompositions of $\om$ are of
direct integral type, and the component states will not lie in the same sector.
Such extremal
decompositions are useless.)

The main point is now that  the states occurring in any further decomposition
of a given
macroscopically pure component of the central decomposition of $\ps$ all
coincide on the centre
$\Z(\A)$ of $\A$. In other words, for the classical observables there is no
distinction between the
central and the extremal decomposition.
 In view of our value
attribution rule, the coefficients of the  central decomposition may therefore
still be
interpreted as Born probabilities. The ignorance interpretation of mixed states
is now only
applied to the central decomposition. Conversely, the uniqeness of the central
decomposition as
opposed to the nonuniqeness of the extremal decomposition may be used as an
argument in favour of
our point of view that only classical observables possess values. For states
$\ps$ of the special type
considered in the main text, the extremal and the central decompositions
coincide.}\ll{gen}}.

The statistical character of quantum mechanics entirely derives from the
coefficients $p_i$ in the
unique decomposition of mixed states into pure ones, {\em if this uniqueness
applies}. Thus it fully
sits inside the user interface. We find this most satisfactory, for it  shows
that probability only
emerges if some part of the system is left out of consideration, and therefore
has a similar
origin as in classical physics: it reflects ignorance   of the discarded part.
 The kernel
of quantum mechanics remains fully deterministic.

 Carefully note, that this does not lead to a
conventional hidden variables interpretation of quantum mechanics, for the part
of the world that is
declared irrelevant is  a quantum system. Also, the elimination of the
irrelevant
observables should not be confused with incorrect attempts to resolve the
measurement problem by
finding the apparatus in a mixed state (namely a reduced density matrix)
obtained by tracing out
either the measured system or the environment; such mixed states do not have a
unique
extremal decomposition, and even if their orthogonal decomposition is unique,
they do not admit a
consistent ignorance interpretation (on the latter point see e.g.\ Beltrametti
and Cassinelli
(1981), van Fraassen (1991), Busch et al.\ (1991)).

If this approach can be made to work, it obviates the need for a `collapse of
the wavefunction' as a
consequence of the act of perception. Recall that in his attempt to resolve the
measurement problem
von Neumann (1932)   introduced an irreversible time-evolution in quantum
mechanics, on top of the reversible unitary evolution given by the
Schr\"{o}dinger equation.
This was motivated by his idea that these two evolutions would  correspond to
the system evolving either
while being perceived, or autonomously, respectively. In contrast, in
approaches based on
superselection rules there is a single time-evolution, and the duality
solecistically identified by von
Neumann actually corresponds to the system being purely quantal, or (partly)
classical, depending
on its algebra of observables. The `collapse' does enter through the back door,
but plays a
different, mind-independent role: as we shall see, it sneaks in through the
definition of the
time-evolution of the algebra of observables of the environment. We stress that
this has nothing to
do with acts of measurement of observation.

The main program in this approach is evidently to firstly describe and justify
the truncations
allowing the ensuing algebra of observables to have a nontrivial centre, and
secondly to derive the
appropriate mixed `post-measurement state' $\psi$ having the desired unique
extremal decomposition.
This leads to considerable difficulties, to which we shall turn first.
\section{A simple model}
The following model of a measurement apparatus was considered by Hepp (also
cf.\ Bona
(1980), Bub (1988)). It seems that most points of philosophical interest can be
discussed in its
context. The apparatus consists of an infinite chain of spin 1/2 particles.
The Hilbert space of states $\H$ (which is  non-separable) is the so-called
complete\fn{This means
that there are no restrictions on the infinite tail. An example of an
incomplete infinite tensor
product would be the Hilbert space completion of the set of those states in
which only a finite
numbers of spin are up. See von Neumann (1938).}
 infinite tensor product of  all single-particle Hilbert spaces ${\Bbb C}^2$.
The
set of all beables would correspond to the algebra $\B(\H)$ of all bounded
self-adjoint linear
operators on $\H$.
 The dynamics of the system is given by  a nearest-neighbour coupling
of ferromagnetic type. Hence the ground state is doubly degenerate: either all
spins are up, or they
are all down.
Equivalently, one may regard this apparatus as a photo-emulsion (which is what
Hepp did).
In this interpretation the `spin up'  state of each spin 1/2 particle is
replaced by an AgBr
molecule (assumed to have just one state of interest), while the `spin down'
state corresponds to an
Ag atom.

The crucial assumption is now that pointer {\em observables} are generated by
local operators of the
type $A_1\ot A_2\ot\ldots A_n\ldots {\Bbb I}\ldots {\Bbb I}\ldots$, in which
only a finite number of
entries $A_i$ (which are 2x2 matrices) are different from the unit matrix $\Bbb
I$.
This assumption says that the algebra $\A$ of pointer observables is
insensitive to correlations at
infinity\fn{$\A$ is given by the weak closure of the linear span of all
operators of the given
type. In this example, the commutant $\A'$ is not abelian, cf.\ footnote
\ref{gen}. We therefore
have to check explicitly whether a given vector $\Psi\in\H$ defines a state on
$\A$ whose extremal
decompoition is unique.}.

On this assumption, $\H$ splits up into disjoint superselection sectors. For
example, the ground
states $\Up$ in which all spins are up, and $\Do$ in which they are all down,
have the property $(A\Up,\Do)=0$ for all $A$ in $\A$. Any two such vectors in
$\H$ between which all
matrix elements of $\A$ vanish, are accordingly said to lie in different
superselection sectors.
Conversely, two vectors which differ only in a finite number of single-spin
states are clearly in
the same sector. If we form a sum $\Psi=a\Psi_1+b\Psi_2$ of two vectors lying
in different sectors,
then  the corresponding state $\ps$ on $\A$ equals $|a|^2\ps_1+|b|^2\ps_2$,
that is, {\em it is
mixed}.

 Now consider the operator $s_n ={\Bbb I}\ot\ldots \sg_3\ot {\Bbb I}\ldots$,
which
has $\sg_3={\rm diag}(1,-1)$ in the n-th entry, and unit matrices everywhere
else.
 From the $s_n$ we build $S=\lim_{N\raw\infty}1/N\,\sum_{n=1}^N s_n$. (This
limit exists in the weak
operator topology and   lies in $\A$.) The operator $S$ has the remarkable
property that its matrix
elements between any two vectors in a given sector coincide, so that it only
`sees' in which sector
a given vector lies. For example, $(S\Psi,\Phi)=1$ for any two vectors
$\Psi,\Phi$ which lie in the same sector
as $\Up$, and $(S\Psi,\Phi)=-1$ in the sector of $\Do$.
Indeed, the  operator $S$ is a classical observable, as defined in the
Introduction.

To apply this setting to the measurement problem, consider a single spin 1/2
particle, whose spin is
measured by the pointer. Assuming that the initial state of the pointer is
$\Up$, this implies
that the post-measurement state of the pointer should be in the sector of $\Up$
in case  the
particle spin is up, whereas it should be in the sector of $\Do$ in the
opposite case.
If so, we can simply look at the operator $S$ to see what the spin of the
particle was.
If the particle is in a superposition $a\up+b\dow$ of up and down states, the
pointer will not be in
the corresponding superposition of $\Up$ and $\Do$: instead, its state $\psi$
will be mixed on $\A$,
and the unique decomposition into pure states is (with some abuse of notation)
$\ps=|a|^2
\Up+|b|^2\Do$.

Hence, as promised, the Born probabilities normally associated with the
particle whose
spin is being measured, emerge as coefficients in the decomposition of the
mixed post-measurement
state of the apparatus. Moreover, the usual measurement `paradox' of arriving
at
never-observed superpositions of    pointer states has been obviated\fn{Cf.\
Dieks (1991) and
Landsman (1991) for two explanations why this `paradox' is spurious anyway.}.
If the post-measurement
emergence of the pointer in a definite state, in which $S$ has either the value
1 or -1, is taken to
be an event, then the occurrence of such an event is definitely predicted,
although which   of the
two possibilities is realized is a probabilistic affair, as usual.  This
statement relies on the
ignorance interpretation of mixed states   allowing a unique extremal
decomposition.

But how is the desired post-measurement state to be arrived at? The problem is
that the initial
pointer state $\Up$ should stay in the same sector if coupled to the state
$\up$ of the particle,
whereas it should evolve into the $\Do$ sector if coupled to the $\dow$ state.
Hepp showed, in models
as well as in a theorem,  that this cannot be done in finite time with an
automorphic
evolution\fn{This is the algebraic counterpart of a unitary evolution in the
usual formalism.
Also cf.\ Landsman (1991) for a further discussion. The
claim
in  Bub (1988) that this can be accomplished nonetheless is too hasty. It is
trivial to find a unitary
group $U_t$ which maps any sector in any other one in a finite time; the point
is that in addition  a
corresponding Heisenberg picture time-evolution of the algebra of observables
should be defined. This
is not done in  Bub (1988); in particular, for his $U_t$ the map
$A(t)=U_tAU_t^*$ maps $A$ outside the
algebra of observables $\A$.  To get back into $\A$, one would need a
conditional expectation
from $\B(\H)$ to $\A$, as in section 4 below. In the present case (and similar
ones) such an object
is not readily available. \ll{bub}}. Heuristically, what has to happen is that
the particle in the
spin down state should flip the first spin in the pointer, then the
second\ldots, so that it is
obvious that one needs an infinitely long time to complete the measurement.
 In other words, the dilemma is this: superselection sectors are defined by the
property that no observable (such as the Hamiltonian) can interpolate between
them. Yet the little
spin which is measured is supposed to cause precisely such a transition.

The best one can achieve is that
$$\lim_{t\raw\infty}(A(t)\Up\ot\dow,\Up\ot\dow)=(A(0)\Do\ot\dow,\Do\ot\dow)$$
for each fixed $A\in\A$;
here $A(t)$ is the Heisenberg picture time evolution of the operator $A$.
This led to (a revival of) the idea, that one ought to regard a measurement as
a process analogous
to a scattering event, which  officially takes infinite time to be completed,
too (in the sense that
the particles only approach their on-shell states in the $t\raw\infty$ limit),
with no one ever
complaining about the idealization this implies.
\section{The disaster of infinite measurement time}
 Unfortunately, one cannot help feeling
uncomfortable with this approach to the measurement problem.
States in different superselection sectors may be thought of as being
`macroscopically' different:
they differ by an infinite number of  single spin states in the pointer. If one
only monitors a
finite number of spins, one cannot detect in which sector the pointer state
lies. We are supposed
to see instantly what the (final) macroscopic state of the pointer is, for
otherwise we would not
accept it as a measurement device.
However, at any finite time this macroscopic state coincides with the initial
state (up to a finite
number of spins). (Note, that  our
alleged ability to see an infinite number of spins as such does not conflict
with the assumption that
the pointer observables are (quasi-) local.  This ability means, that we can
monitor observables like
$S$. A nonlocal observable, however, would detect the quantum interference
between states in different
sectors, that is, such an observable would correlate an infinite number of
spins. And to see such
correlations is surely beyond us).

The achievement of a limit means, that for finite but very large time the
quantity that is to have
the given limit (namely the matrix element of a fixed local observable $A$),
differs from its limit
value by an arbitrarily small number. Since in any finite time only a finite
number of spins in the
pointer have flipped, that is, the pointer is still in its original sector
$\Up$, we can only have the
illusion that the measurement has almost been completed because the fixed
operator $A$ above monitors
a finite number of spins (or, as in the case of the pointer observable $S$,  is
a weak limit of such
operators). But it is precisely the infinite `tail' of the pointer which
determines in which sector it
is, and which `macroscopic' properties it has. If we really `observe' the
entire pointer, we would at
any finite time conclude that the pointer is still in the $\Up$ sector, and
abandon the hope that any
form of measurement of the particle spin is being performed.

A somewhat confusing variant of  this argument  was forwarded in  Bell (1975),
where the poverty of the way the infinite-time limit of the measurement is
approached is illustrated
by the following observation. Suppose we give the operator $A$ an explicit
time-dependence $A_t$,
which exactly cancels its Heisenberg-picture time-dependence $A(t)$. Then
$(A_t(t)\Up\ot\dow,\Up\ot\dow)$
 is obviously time-independent, and no limit will ever be reached! Of course,
we
previously assumed $A$ to be fixed, and by replacing it with $A_t$ one inserts
an infinite
family  of operators. To interpret this argument properly, it should be
realized
that it can be forwarded against the idea  of measurement altogether, in that
the
explicit time-evolution $A_t$ simply `undoes' the measurement. Thus we briefly
discuss
this possibility.

The `undoing' of quantum measurements was discussed by Peres (1980, 1986), who
claimed its
impossibility. The argument is that the construction of the  explicit
time-dependence $A_t$ amounts
to an inversion of the equations of motion; this would imply that given the
pointer state at time
$t$,    Bell (1975)  would have to calculate its state at $t=0$. For the
Heisenberg
picture matrix element $(A_t(t)\Psi_1,\Psi_2)$ equals the Schr\"{o}dinger
picture
$(A_t\Psi_1(t),\Psi_2(t))$, which indeed by construction equals
$(A\Psi_1(0),\Psi_2(0))$. If the
pointer is sufficiently large (it is, indeed, assumed to be infinite), that
would certainly have been
beyond Bell's abilities, were it not for the fact that the initial pointer
state is known to be $\Up$!

Thus Peres' counterargument seems to lose its weight. On the other hand, one
might argue that the
initial pointer state does not need to be exactly $\Up$; it suffices for it to
be in the same
superselection sector. But to use the type of complexity argument in Peres
(1980), a sufficiently
large number of pointer spins should be randomly determined in the initial
state. But how large is
`sufficiently large', compared to the infinite number of spins that are fixed
to be up in this
sector? Fortunately, irreversibility arguments attempting to make the
undoing of the measurement impossible by declaring  the initial state to be
randomish are very
powerful indeed, but need a little extra ingredient, as we shall see in the
next section.

A further point is that one should acknowledge that all pointers are actually
composed of a finite
number of particles\fn{The fact that the stuff of the world is quantum fields
rather than particles
does not affect this discussion, despite the fact that a field theory has an
infinite number of
degrees of freedom even in a localized region. For the local algebras in
quantum field theory are
factorial von Neumann algebras of type I (in the non-relativistic case) or type
III (in the
relativistic case), and these admit only one (normal) representation up to
unitary equivalence.
Hence they cannot lead to superselection rules.}; the
infinite system is the `exact background theory' that B-realists always call
for. But to what extent
does the theory of a finite pointer approximate this exact theory? One usually
hears that on account
of the Stone-von Neumann uniqueness theorem, superselection rules do not exist
in finite systems.
Hence one would have to go for approximate superselection rules, in the sense
that one tries to find
classes of states with the property that matrix elements of `relevant'
observables between any two
vectors lying in a different class are `small', cf.\ Lloyd (1988). The idea
here is that only a
relatively small number of spins determine the  relevant algebra of
observables, the much larger
remainder playing the role of the infinite tail of the exact theory.

This is problematic; for one thing, the ignorance interpretation of mixed
states only works if
some unique decomposition is applied (such as the central decomposition, cf.\
footnote \ref{gen}).
But if the superselection rules are only approximate, the centre is trivial and
one does not even
have  suitable ingredients for a decomposition of an exact mixture, let alone
of an approximate
one.
 And what is small and what is large? The relevant
scale should be set externally.
It all starts to sound pretty vague. (This dilemma has led Bub (1988)
  to propose that one needs truly infinite apparatuses to resolve the problem;
hence only
infinite systems can have `objective' properties.  While we are sympathetic to
the idea of an
approximate reality in finite systems, the problem of the infinite-time limit
is not resolved in this
way (cf.\  footnote \ref{bub} on  Bub (1988)). Moreover, the origin of the
objectification of
properties of infinite systems still lies in the choice of local observables;
this idea, however, is
much more convincingly implemented if one invokes the environment (see below).)

Apart from this conflict between macroscopic observability and locality of the
interaction with the
measured system, one may argue that resolutions of the measurement problem
involving superselection
rules of the apparatus are incomplete even if they work, for the `algebra of
observables' is not an
intrinsic object: by name alone, it presupposes a theory of the observer and
his/her/its coupling to
the apparatus. Hence in any case the choice of the observables, and thereby the
superselection rules,
must be motivated externally. So far, their introduction has (at best) merely
parametrized the
classical behaviour.
\section{Environmental superselection rules}
Let us return to a finite (but large) pointer, into whose state space we still
would like to
introduce superselection sectors in some sense. A crucial assumption in  the
Stone-von Neumann
uniqueness theorem, which is usually quoted as excluding superselection rules
in finite systems, is
that one has a simple algebra\fn{A simple $C^*$-algebra is one without closed
2-sided ideals. Since
the kernel of any representation forms an ideal, it follows that all
representations of simple
algebras are faithful.}. Indeed, a finite system does have superselection rules
if its algebra of
observables is not simple,  cf.\ Landsman (1991). To illustrate this point,
consider the algebra
$M_2$ of $2\times 2$ matrices, acting on the Hilbert space ${\Bbb C}^2$. Let
the basis vector ${\bf
e}_1$  stand for a photon\fn{Any light particle would do. We ignore helicity,
anyway.} state
localized very far away from the photon state ${\bf e}_2$. Saying that $M_2$ is
the algebra of
observables of this system amounts to pretending that an operation exists by
which one can determine
quantum interference between these basis states. This not being feasible\fn{
{\sp The claim in
Aharonov et al.\ (1986) that nonlocal states can sometimes be measured is of no
relevance here. To
determine  quantum interference between localized states $\bf x$ and $\bf y$
one needs an operator
$A$ such that $(A{\bf x},{\bf y})\neq 0$. But the observables considered in
Aharonov et al.\ (1986)
are of the form $A_1+A_2$, where  the localization region $\O_1$ of $A_1$
contains $x$ but not $y$,
and {\em vice versa} for $A_2$. Such operators have vanishing matrix elements
between $\bf x$ and
$\bf y$. Moreover, they are not nonlocal at all in the usual sense (Haag,
1992): they are localized in
$\O_1\cup \O_2$.}},
 one
will have to admit that the actual `effective' algebra of observables, relevant
to a local observer
who/which is unable to perform highly nonlocal measurements, is $D_2\simeq
{\Bbb C}\oplus {\Bbb C}$,
the algebra of diagonal $2\times 2$ matrices. This surely has two inequivalent
representations\fn{The first one maps the first copy of $\Bbb C$ to itself, and
the second copy to
zero; the second one does the opposite.}, and hence two superselection sectors.
The first sector
has ${\bf e}_1$ as its only normalized vector (up to a phase), and the second
sector contains merely
${\bf e}_2$.  This is possible, as $D_2$ is not simple; instead, it is the
direct sum of 2 copies of the algebra of the complex numbers.

Our finite pointer can inherit these superselection rules in the following way.
If the pointer
contains $N$ spins, its algebra of observables is $M_{2^N}$. The algebra of the
total system pointer + photon is then $\A_{\rm tot}=M_{2^N}\ot D_2$, and has 2
sectors.   In a
first attempt to exploit this, one would like to find an initial photon state
$I_E$ and a
pointer-photon interaction such that,  any   pointer state $U$ in which an
overwhelming majority of the spins is up, coupled to (i.e., tensored with)
$I_E$ evolves into
$U\ot {\bf e}_1$, whereas the analogous states $D\ot I_E$ with most spins down
evolve  into
$D\ot {\bf e}_2$.  These two final states are in different superselection
sectors, and the same
effect has been achieved as in the infinite pointer. The role of the infinite
tail of the latter is
now played by the photon, which flies away to the Andromeda nebula. (As in the
preceding section,
this transition cannot be achieved with an automorphic time evolution; see
below for the resolution.)

 If we now re-introduce the
particle whose  spin is measured, we see that once again a superposition $a
\up+b\dow$ of states of this
particle does not lead to a corresponding superposition of the pointer +
photon, but to a mixture
whose decomposition into pure states has the Born numbers $|a|^2$ and $|b|^2$
as coefficients.
The origin of this `collapse of the wavepacket' now lies in the relative
delocalization of the
photon states ${\bf e}_1$ and ${\bf e}_2$, combined with the presumed locality
of the observer,
rather than in the locality of the pointer observables (see below for a more
detailed explanation of
this point). Also, since the pointer is finite, one only needs a finite time
for all the spins in  it
to flip if necessary, and the photon flies away rather quickly, too.

Unitarity of the time evolution implies that the initial photon state has to be
fixed in order to
achieve this scenario. This is undesirable, firstly because photons are not
part of the construction
of pointers (they belong to the `environment'), and secondly because the
invariability of the
initial state leads to the possibility of `undoing' the measurement, as
explained earlier.
The way out is to consider a very large environment $E$ (perhaps consisting of
an enormous number of
photons). Its algebra of observables $\A_E$ is still supposed to lead to
superselection rules on its
state space, by not containing   operators whose measurement would involve the
detection of quantum
interference  between states which are localized lightseconds or more away from
each other. The
mechanism by which the omission of such operators from the original simple
algebra leads to the ensuing
algebra being non-simple, is exactly the same as in the example above.

The general strategy to obtain superselection rules (caused by the locality of
the observer) is to
identify subsystems between which no correlations are observed. For example,
one may imagine a
sphere with radius $R$ around the observer, and stipulate that no correlations
with
any region outside this sphere are observed\fn{This differs from the approach
of Wan and Jackson
(1984). Their condition for an operator $A$ on $L^2({\Bbb R}^3)$ to be an
observable is that
$(A\Psi,\Psi)=0$ for all $\Psi$ localized outside the sphere. Our condition is
that
$(A\Psi,\Phi)=0$ for all $\Psi$ localized outside the sphere {\em and}
$\Psi,\Phi$ having disjoint
(essential)  support.}. The precise value of $R$ is rather arbitrary, as long
as it is  very large\fn{
Sending $R$ to infinity is not the same as having no dismissed correlations at
all; it rather
corresponds to ignoring correlations at infinity. The superselection rules of
local field theories
are a consequence of this.} compared to the size of the system studied. A
measurement is then
completed the moment the objects (e.g., photons) carrying the quantum
correlations of the system
after the interaction with the environement have left the sphere.

The only requirement on the dynamics is that $U\ot I_E$ and $D \ot I_E$ evolve
into $U'\ot e_n$ and
$D'\ot e_m$, respectively, where the  final environment states $e_n$ and $e_m$
(which may vary with
the initial environment state
$I_E$) lie in different superselection sectors, and $U'$  and $D'$ are `close'
to $U$ and $D$ (in a
sense to be made more precise shortly). If the interaction between pointer and
photon bath is
suitable (see below), this  condition will be satisifed by the overwhelming
majority of initial
states, for the environment has a large number of particles, each of which
interacts with the
pointer, and it suffices if merely one of these particles has an initial state
leading to orthogonal
final states after coupling with the pointer states $U$ and $D$, respectively.
 Even if that is not the case, the phases of the terms in the inner
product $(e_n,e_m)$ (which is a sum of products of a gigantic number of
single-particle inner
products) will be random so that the product vanishes (a point forcefully made
by van Kampen (1988),
who added that someone who does not accept this doesn't understand what physics
is).
Any reasonable assumption on what the   algebra of observables of the
environment relative to a
local observer is will then imply that $e_n$ and $e_m$ are not merely
orthogonal but lie in
different sectors on top of that.

 The technical conditions under which the above scenario works were
established by Zurek (1981, 1982, 1993)  (also cf.\ Joos and Zeh (1985) and Zeh
(1970)), where the
idea of introducing the environment in the measurement problem was first
proposed).   In these
papers, the central issue is to identify the family of pointer states
$\{p_n\}_n$ which has the
property we ascribed to $U$ and $D$, that is, that for most environment states
$I_E$,  $p_n\ot I_E$
evolves into $p_n\ot e_n$ for some environmental state $e_n$, with $e_n$ and
$e_m$ orthogonal for
$n\neq m$.  This is so if the operator $P$ which has the $p_n$ as eigenstates
commutes with the
interaction Hamiltonian describing the coupling between the pointer and the
environment.

As a first application of this rule, we can easily see why the presumed
locality of an observer
 excludes the possibility of measuring correlations between two photon states
which are
delocalized relative to each other. For let ${\bf e}_1$  and  ${\bf e}_2$ have
their previous
meaning.
To detect interference between these states, one would need to bring
eigenstates of the Pauli
matrices $\sg_1$ or $\sg_2$ into correlation with the eigenstates of a certain
operator relevant to
the observer; the former eigenstates here play the role of the $p_n$. But a
local observer-photon
interaction Hamiltonian must have the form $H_I=[{\bf e}_1] \Phi_1 + [{\bf
e}_2]\Phi_2$, where
$[{\bf e}_i]$ is the projector on ${\bf e}_i$, and the $\Phi_i$ are operators
on the state space of
the observer. Now clearly the commutators $[H_I, \sg_1]$ and $[H_I,\sg_2]$
are  nonzero, so it follows that the establishment of the desired nonlocal
correlation must involve a
nonlocal interaction Hamiltonian (an obvious result!).

If the pointer is large, one can relax the stability condition on $p_n$ under
time-evolution,
so that one is in an even more comfortable position than Zurek's. For in that
case the only
requirement is that $p_n\ot I_E$ evolves into $p_n'\ot e_n$, where $p_n'$  is
macroscopically close
to $p_n$ (and $(e_n,e_m)\simeq \dl_{nm}$, as before). To illustrate what this
means, consider the
case where the pointer observable $P$ is taken to be $ S= 1/N\,\sum_{n=1}^N
s_n$, defined as in the
previous section, but not taking the limit $N\raw\infty$. The eigenstates of
$S$ are states of the
form $\ot_{n=1}^N\, u_n$, where each $u_n$ is an eigenstate of $\sg_3$. The
spectrum of $S$ is
$\{(2N_+-N)/N\:\, |N_+=0,\ldots,N\}$; here $N_+$ is simply the number of
$u_n$'s whose spin is up.
This spectrum is highly degenerate, and, more importantly, for large $N$ the
eigenvalues are closely
packed. To read off the measurement outcome it is sufficient that the
expectation value of $S$ in the
final state is either close to 1, or to -1. So in the above argument, $p_n$ and
$p_n'$ should have
almost the same expectation value of $S$. This modified requirement makes
Zurek's stability condition
far more robust, and easier to satisfy.

The transition $p_n\ot I_E\raw p_n\ot e_n$ is evidently a time-dependent
process, so let us write
$e_n(t)$ to make this clear. It turns out that $(e_n(t),e_m(t))\simeq \dl_{nm}$
only for $t\raw
\infty$ and an infinite environment. For finite environments the inner product
typically becomes
exponentially and phenomenally small quickly, gets ever smaller, until it
returns to sizable values
after supercosmic timescales (Poincar\'{e} recurrence), cf.\  Zurek (1981,
1982, 1993), Joos and Zeh
(1985), and refs.\ therein. The consequences of this behaviour may be
illustrated on our familiar
one-photon example. So let the post-measurement state vector be $a\, U\ot ({\bf
e}_1+\ep(t){\bf e}_2)
+ b\, D\ot ({\bf e}_2+\ep(t){\bf e}_1)$; we ignore terms of order $\ep^2$, so
that this state is
normalized. Under  the same assumptions as before, this state is mixed, and the
corresponding density
matrix has the unique extremal decomposition $\rh(t)=|a|^2 [v_1]
+|b|^2[v_2]$, where $[v_i]$ denotes the projector on $v_i$,
$v_1=|a|^{-1}( a\, U+\ep(t)b\, D)\ot {\bf e}_1$ and  $v_2= |b|^{-1}( b\, D
+\ep(t)a\, U)\ot {\bf
e}_2$.
  Since the decomposition is unique, we are still entitled to apply the
ignorance interpretation of
such mixed states, which implies that the coupled system is either in the state
$v_1$ (with
probability $|a|^2$), or in the state $v_2$ (with probability $|b|^2$). Here
$\ep(t)$ is
very small,
 and gets increasingly smaller, as indicated above.

We see that no difficulty arises. All macroscopic observables (like $S$) will
equate $v_1$ with
$U\ot {\bf e}_1$ and $v_2$ with $D\ot {\bf e}_2$ (up to terms of order
$\ep^2$). To distinguish
between $v_1$ and  $U\ot {\bf e}_1$, one would have to perform interference
experiments involving
all spins in the chain, and even so one would detect an effect of order
$\ep(t)$, which, as
mentioned before, is astoundingly small. In case one is actually able to do all
this, it has to be
concluded that the pointer has failed its service as a measurement apparatus:
if an undesired
superposition can be detected, one should stop looking for arguments why it is
actually not there.
But if it is there, the measurement problem does not arise, for no measurement
has been performed in
that case!

The same would-be difficulty arises in the modal interpretation (see below),
and, as pointed out by
Dieks (1994a), in practically any measurement theory. His answer is as
satisfactory as the one given
above; he simply points out that in the modal interpretation the pointer
possesses a (macroscopic)
spin in the direction determined by $ a\, U+\ep(t)b\, D$; this direction is
practically
indistinguishable from the $z$-axis, and the problem has been obviated (the
second point discussed by
Dieks concerns the situation where $a\simeq b$, but the difficulties this leads
to are only relevant
for small pointers)\fn{Also cf.\ Bacciagaluppi and Hemmo (1994) for an
extensive discussion of this
controversial issue.}.

 The argument above was given in the Schr\"{o}dinger picture, where the states
evolve in time.
This conceals a very important aspect of our scenario.  In the single-photon
example, where the
algebra of observables was taken to be $D_2\subset M_2$, the time-evolution of
the state vectors
$\Psi\in {\Bbb C}^2$ is simply given by a unitary one-parameter group $U_t$ in
$M_2$, i.e.,
$\Psi(t)=U_t\Psi$. This defines an automorphic time-evolution on $M_2$ by
passing to the Heisenberg
picture, in which $A(t)=U_t^* AU_t$ for each  $A\in M_2$. However, this will
generally map elements of
$D_2$ outside this algebra, unless each $U_t$ itself lies in $D_2$. But  if
 the initial state is $c_1 {\bf e}_1 + c_2 {\bf e}_2$, then such a
time-evolution would merely change
the phases of the $c_i$, and the whole show would have to be canceled. However,
the observed
time-evolution $A_{\rm obs}(t)$ of an observable $A\in D_2$ is obviously not
$A(t)$, but merely the
part of this operator which lies in $D_2$, that is, its diagonal. Defining the
projectors
$P_i=[{\bf e}_i]$ ($i=1,2$) we can write this as  $A_{\rm obs}(t)=\sum_i P_i
A(t)P_i$.

The proper setting for this discussion is the algebra of observables $\A$ of
the system, the
apparatus, and the environment together. In our context its superselection
structure only derives from
the last component, but this particularity is irrelevant for the following
construction.
Let $\H$ be the Hilbert space of state vectors of $\A$; according to our
assumptions, this contains
a number of superselection sectors $\H_i$ (for simplicity we assume discrete
superselection rules).
We   denote the projector onto $\H_i$ by $P_i$.
If $U_t$ is the unitary one-parameter group defining time-evolution on $\H$,
then, as before, one
has\fn{Technically, the map $E(A)= \sum_i P_i  AP_i$  defines a conditional
expectation
$E:\B(\H)\raw \A$. See Davies (1976) and Landsman (1991) for other uses of such
maps in measurement
theory.}
 $A_{\rm
obs}(t)=\sum_i P_i U_t^* AU_t P_i$. For $U_t\notin \A$ this evolution is
non-automorphic and
irreversible, and the origin of this irreversibility (namely projection on the
relevant degrees of
freedom) is the same as in conscientious non-equilibrium statistical mechanics,
 e.g.\ Balian et al.\ (1986).

Also, one is immediately reminded of the collapse of the wavefunction. For if
$\rh$ is a density
matrix then the expectation value ${\rm Tr} \rh A_{\rm obs}(t)$ equals ${\rm
Tr} \rh_{\rm collapsed}
A(t)$, where $\rh_{\rm collapsed}=\sum_i P_i \rh P_i$. But there are two
important differences between
the above `collapse' of $\rh$ to $\rh_{\rm collapsed}$, and the traditional
(von Neumann) collapse.
Firstly, the former does not take place instantaneously; indeed, it has nothing
to do with
perception and little with measurement. Rather, it is a consequence of the
selection of the algebra
of observables. Secondly, the collapse does not take place with respect to the
spectral projections
of the system observable that is measured, but with respect to the projectors
on the superselection
sectors. Since in our approach the latter are neither related to the system
that is measured, nor to
the measurement apparatus, we see that the two collapses should not be confused
indeed\fn{Though if
one interprets $\A$ as the algebra of system observables then  formally
$\rh_{\rm collapsed}$ is what
one obtains from L\"{u}ders' rule, applied to the measurement of central
elements of $\A$
(whose spectrum is highly degenerate if $\A$ is sufficiently
non-commutative).}.

A further point concerns the argument why the pointer should be in either the
state $U$ or in $D$,
given that the total system of pointer + environment has the state vector $a\,
U\ot {\bf e_1}+ b\,
D\ot {\bf e}_2$. Zurek here mumbles something about ``the right of a
macroscopic (but ultimately
quantum) system to be in its own state''  (Zurek, 1993, p.\ 287), and proceeds
with the usual argument
of reducing the density matrix of the total system by tracing over the
environment. This leaves the
mixed state $|a|^2 [U]+|b|^2 [D]$ of the pointer, which is then construed
according to the
ignorance interpretation. The last step is justified by Zurek by the idea that
the basis relative to
which this reduced state is decomposed is precisely the preferred pointer basis
(Zurek, 1981).

Although this reasoning leads to the correct result, it is based on a
questionable application of
the ignorance interpretation (namely to mixtures whose extremal decomposition
is non-unique).
 Moreover,
it obscures the origin of the `factualization' of the states $U$ and $D$, viz.\
the assumed
superselection rule of the environment, which itself is caused by the
delocalization of the quantum
coherence (which according to the `kernel' of the theory is always there) and
the locality of the
observer.  Hence in our opinion the correct argument is as follows. Without
superselection,
i.e., serious restrictions on what the observables of the theory are, there is
no collapse of the
wavepacket.    On the other hand, if the operators interpolating
between highly delocalized states of the environment are removed,
the state $[a\, U\ot {\bf e_1}+ b\, D\ot {\bf e}_2]$ collapses to
$|a|^2 [U\ot {\bf e}_1]+|b|^2 [D\ot {\bf e}_2]$ in the case at hand. Thus one
not only has
the collapse of a pure to a mixed state, but on top of that the mixed state has
a unique decomposition
into the pure states $U\ot {\bf e}_1$ and $D\ot {\bf e}_2$. These are product
states, hence the only
state of each subsystem consistent with each of these pure states on the
combined system is the one
occurring as a factor in the tensor product.

However, this does not mean that the pointer observable $S$ now has a value, at
least not in our
interpretation. In the Introduction we mentioned that in our interpretation an
observable can only
have a given value (in other words, possess a property) if it is in the centre
of the algebra of
observables; in particular, we abandon the usual stipulation that an arbitrary
observable possesses a
value (namely, the pertinent eigenvalue) if the system is in an eigenstate of
the corresponding
operator. Hence, if we take the (finite) pointer by  itself, we are unable to
conclude that the
macroscopic spin observable $S$ has the value $+1$ even if all the spins in the
pointer are up.
Indeed, since the algebra of observables of the  finite pointer has no
superselection rules, none of
its observables has a value in any state: a pure quantum system simply has no
properties in the usual
sense.

Our illusion that the pointer is `up' is entirely caused by its photon
environment. It suffices to
illustrate this matter  in the model  we used
before, where the environment consists of a single photon. The total
algebra of observables was $\A_{\rm tot}=M_{2^N}\ot D_2$; its centre is simply
${\Bbb I}_{2^N}\ot
D_2$, where ${\Bbb I}_{2^N}$ is the identity operator for the pointer. In
particular, the photon
operator ${\Bbb I}_{2^N}\ot \sg_3$ is in this centre, and in our interpretation
we are allowed to
conclude that it actually has the value $+1$ in any state of the form $\Psi\ot
{\bf e}_1$, where
$\Psi$ is an arbitrary pointer state. Given the dynamics of the
pointer-environment interaction,
which brings the states $\Up$ and ${\bf e}_1$ in correlation, we may then
conclude that the pointer
is in the state $\Up$ if we observe the photon in the state ${\bf e}_1$, and,
by implication, that
the little particle whose spin was measured by the pointer, is in the state
`spin up' as well. {\em
Yet none of the pointer observables possess values corresponding to this state,
because they do not
possess any value at all}. This is an entirely satisfactory conclusion, for all
we actually {\em see} is the photon environment. Of course, this conclusion is
really convincing only
if the photon environment is huge; the qualitative argument is the same also in
that case. And when
means of monitoring other than vision apply, an analogous argument hawks about,
the photon environment
replaced by whatever is being monitored by the observer.

Finally, note that the pointer itself does have properties if it is infinite,
and its algebra of
observables has a nontrivial center. The point of the discussion in section 3
is that this property
does not yet suffice to make it a measurement apparatus.
 \section{Some remarks on the modal interpretation}
Many issues concerning environment and superselection are also relevant for the
modal
interpretation of quantum mechanics. This programme is still under development;
there are four
different versions, due to    van
Fraassen (1991), Kochen (1985),  Healey (1989),
and Dieks (1994a)\fn{We have cited the most accessible and relevant
publications. There are older
papers by van Fraassen and by Dieks on this subject, and the four authors have
developed their ideas
independently.}.
The essential business of the modal interpretation is to put a subtle form of
the projection
postulate neither in the mathematical formalism, nor in Nature itself, but in
the rules of
interpretation of the formalism.  As such it is able to provide a realistic
interpretation of quantum
mechanics (in the sense of `A-realism', cf.\ the Introduction) without changing
its mathematical
structure at all.  Further using the terminology of the Introduction, the modal
reading of quantum
mechanics  puts the physical interpretation of the formalism into the kernel,
and aims at describing
observations as special instances of properties of systems possessed anyway.

 The modal interpretation maintains the rule that an  observable of the total
system actually
possesses a value if the total  system is in an eigenstate of it. However, the
eigenvector-eigenvalue
link is  dropped as a bidirectional connection, in that observables of
subsystems may have values in
states where the von Neumann liturgy would deny that these values are
possessed.

The starting point of the modal interpretation (at least in the version of
Dieks (1994a,b), on
which we will concentrate) is an arbitrary\fn{For Healey (1989) there is a
preferred decomposition
corresponding to elementary particles.}  decomposition of the world $S$ into
(say) two parts $S_1,\,
S_2$, so that (in obvious notation) $\H=\H_1\ot \H_2$, and the state of the
whole system $\Psi\in\H$
has the Schmidt decomposition $\Psi=\sum_{i=1}^N c_i \phv_i\ot \chi_i$.
We assume that all the $\phv_i$ and $\chi_i$ are normalized, so that
$\sum_i|c_i|^2=1$; $N$ may be
infinite, and  we take $\H$  separable. Let us now concentrate on the subsystem
$S_1$ {\em taken by
itself}.
 Given $\Psi$,
the Hilbert space of $S_1$ decomposes as $\H_1=\oplus_{i'=0}^{N'} \H_1^{(i')}$.
Here firstly
$\H_1^{(0)}$ is the subspace orthogonal to all $\phv_i$, secondly $\H_1^{(i')}$
equals the one-dimensional space spanned by $\phv_i$  whenever $c_i$ is
nondegenerate, and finally
all
 vectors $\phv_i$ with identical coefficients $c_i$ together span a single
summand $\H_1^{(i')}$.
Thus apart from $0$ the index $i'$ takes the same values as $i$, except that
those values of $i$
which correspond to the same numerical value of $c_i$ are combined into a
single $i'$.

The crucial step in the modal interpretation is to specify which observables in
$\B(\H_1)$ possess
values, given the total state $\Psi$. In a slight modification of the
literature\fn{The choices in
Dieks (1994a), Dieks (1994b), and Bacciagaluppi and Hemmo (1994) are all
different from each other, as
well as from our prescription. Dieks (1994a) excludes any operator acting
nontrivially on
$\H_1^{(0)}$. Dieks (1994b) includes  all such operators as long as they act
trivially on the
orthogonal complement of $\H_1^{(0)}$, and assigns the possessed value 0 with
probability 1 to them.
The choice of Bacciagaluppi and Hemmo (1994) does not only depend on $\Psi$,
but in addition on the
contingent value state.} we stipulate that each self-adjoint element of the von
Neumann algebra $\A_d$
of all  operators which are diagonal w.r.t.\ the above decomposition  possesses
a value.
An operator $A\in\A_d$ necessarily has discrete spectrum, and has the form
$A=\sum_{i'}a_{i'}P^{(i')}$,
where  $P^{(i')}$ is the orthogonal projector onto   $\H_1^{(i')}$. When all
$a_{i'}$'s are
different\fn{In which case $A$ may still have degenerate spectrum, namely
whenever the Schmidt
decomposition is degenerate.}, the value $a_{i'}$ is possessed with probability
$p_{i'}={\rm dim}\,
\H_1^{(i')} \cdot |c_i|^2$, where $i$ is related to $i'$ as explained in the
previous paragraph. (We
put $p_0\equiv 0$.) If there are degeneracies among the $a_{i'}$  one merely
has to sum over the
$p_{i'}$ corresponding to the degenerate subspaces. Note that
$\sum_{i'}p_{i'}=\sum_i|c_i|^2=1$.
  A similar
assignment of properties holds for $S_2$ taken by itself.

We have written the modal interpretation in the above form in order to stress
the analogy with the
value attribution for systems with superselection rules discussed in this
paper. The point is that
both $\A_d\subset \B(\H_1)$ and $\Z(\A)\subset\A$ are commutative subalgebras
of the algebra of
beables and of observables of the system in question, respectively. It is clear
that at any fixed
time one may consistently assign properties to all observables in such
subalgebras.
In the modal interpretation  this subalgebra depends on the choice of subsystem
and on
the state of the total system. In the superselection approach it originates in
the observer and the
environment. (Note that $\B(\H_1)$ and $\A$ have a rather different structure,
e.g.,
the former does not have a
nontrivial centre.)

The modal interpretation  is tailor-made to add the finishing touch to
the environmental approach in its original formulation (Zurek, 1982, Joos and
Zeh, 1985), that is,
without superselection rules. For, as we mentioned earlier, the aim of this
approach is to interpret
`pointer' states of the form $p_n$, which have the property that they couple to
the environment $E$
according to $p_n\ot I_E\raw p_n\ot e_n$ for `arbitrary' initial states $I_E$
of $E$, and
$(e_n,e_m)\simeq \dl_{nm}$, as classical states. The modal interpretation
provides exactly the
missing link allowing such an interpretation, which helpfully shows that even
such an intuitively
attractive solution to the measurement problem as the  environmental one,
requires a double Dutch
extra interpretative rule of quantum mechanics.

The value attribution given by the modal interpretation is equivalent to  the
ignorance interpretation
of mixed states in the following sense.
If the extremal decomposition of the mixed state obtained by restricting $\Psi$
to (say) $S_1$ is
nonunique (which is the usual situation), the
 modal interpretation precisely leads to the correct interpretation of the
orthogonal decomposition
(which most often is unique, and corresponds to the spectral decomposition of
the reduced density
operator). If, on the other hand,  one of the subsystems (say $S_2$) possesses
superselection rules,
then after a short  time the `pointer basis' of $S_1$ is singled out by the
Schmidt decomposition,
and the modal interpretation ascertains that $S_1$ taken by itself is
effectively in one of the
states\fn{Here meant as a value state in the sense of van Fraassen (1991).}
$p_n$, quite in accordance with our predilection.

However, this is precisely where we feel that the modal interpretation
overshoots its aim: for it
applies whether or not $S_1$  or $S_2$ are
macroscopic, possess superselection rules, etc. For example, the moon may
acquire properties
(that it would not have had otherwise)  by
becoming correlated with a mouse, and this seems almost as bad to us as the
moon only existing when
the mouse looks at it (as in extreme versions of the Copenhagen interpretation,
ridiculed in this way
by Einstein ).
Thus to our mind the modal interpretation   has a similar flaw as B-realists'
approaches to the
measurement problem: it is made regardless of the situation at hand (system
small or large? observer
present and/or relevant? what are the relevant degrees of freedom?), and thus
provides a universal
rule in a problem where the particulars seem to be of prime importance.

In addition, there are two (related) technical difficulties with the modal
interpretation, which are
absent in our approach. The first problem is that of property composition of
subsystems\fn{Healey's
version of the modal interpretation is free of this difficulty, at some other
expenses, cf.\
Bacciagaluppi (1994).}; it appeared in a seminar by R. Clifton and was further
analyzed by
Bacciagaluppi (1994). The  perplexity is that if $S_1$ and $S_2$ possess
property $P_1$ and $P_2$,
respectively, one cannot conclude that the combined system $S_1\& S_2$
possesses the combined
property $P_1\& P_2$. This is even true if one of the $P_i$ is the trivial
property always possessed
(which is represented by the unit operator). Hence an observable $A$ may apply
to $S_1$ but
$S_1\otimes {\Bbb I}$ may then not apply to the total system. Our approach does
not have this
problem, because the centre of an algebra is naturally  contained in the center
of any tensor
product in which the algebra appears as a factor.

The second brain-twister concerns the combination of properties of a single
system at different
times\fn{This is being studied by (at least) D. Albert, D. Dieks, A. Kent, and
P. Vermaas. We learnt
of the difficulty from Kent.}. It amounts to the fact that if the system has
properties $P_i$
at various times $t_i$, then the combination `$P_1$ at $t_1$ {\em and} $P_2$ at
$t_2$ {\em
and}\ldots cannot really be regarded as a property, because the different
(contingent) properties of
that nature do not combine well under the usual rules of probability theory. In
other words,
the properties assigned by the modal interpretation at different times do not
necessarily form a
consistent history in the sense of Omn\`{e}s (1992), Gell-Mann and Hartle
(1990, 1993).

Our proposed value attribution does not suffer from this blemish, for the
projection operators
defining properties are central. Therefore, any history composed out of them
automatically satisfies
the so-called consistency conditions, which guarantee that the joint
probability distribution of the
multi-time properties behaves correctly under taking marginals.
Conversely, this could be taken as an argument in favour of our value
attribution prescription,
which only assigns properties to classical observables.
The price we pay is that (as we have seen) the time-evolution on the algebra of
observables $\A$ is
generally non-automorphic, so that the centre is not necessarily stable under
time-evolution.
Hence a given observable may be classical at one time, but not at another.

Fortunately, the modal interpretation and the superselection approach (both
crucially amended
by environmental ideas; for the former this is explained in Dieks (1994a), for
the latter we refer to
the present paper) completely agree when applied to everyday situations. And it
is precisely the
explanation of common or garden life by quantum mechanics that causes
B-realists such night
thoughts\ldots  The essential point of coincidence between the two is that a
subsystem may have
properties that the total system does not have: objectification is achieved by
ignorance.
The ignorance in the former relates to the step of taking a subsystem by itself
in order to assign
properties to it, whereas in the latter the blindness is to the correlations in
the
discarded part of the world.
\section{Discussion} The bravura language used in this paper is not meant
to conceal the fact that only the barest outline of a resolution (or, better,
dissolution) of the
measurement problem along the lines sketched has been given. More detail was
not omitted for reasons
of space or appropriateness to the journal, but because what has been written
is all there is. What
more is to be done?

Philosophically, the possibility that the abeyance of a certain class of
observables is the origin of
stochasticity in quantum mechanics deserves to be investigated. We stress that
such an account
should not be confused with the  kind of ignorance about the state of some
discarded subsystem (or environment) that is usually invoked.
 Both types of ignorance are codified by restricting a state on some algebra to
a subalgebra; when
the latter is a simple factor in a tensor product this amounts to partial
tracing, and leads to the
usual sort of nescience. In our proposal, on the other hand, one restricts to
an algebra with a
centre, which restriction cannot be described by partial tracing, for the
original algebra is not
the tensor product of the reduced one with some other algebra.

In this light, arguments in favour of the ignorance interpretation of mixed
states (that is,
only those whose extremal decomposition is unique, of course) would be welcome;
such an interpretation
is necessary to comply with the empirical predictions of quantum mechanics, but
is not at all obvious
(cf.\ van Fraassen (1991, sect.\ 7.3)). Similarly, it would be welcome to have
an interpretation of
the notion of a state which goes beyond the minimal definition as a rule to
compute expectation
values of observables. This definition begs the question of how such
expectation values are to be
interpreted, and is only unambiguous for the restriction of a state to the
centre  of an algebra of
observables with superselection rules. For in that case, a state is simply a
value attribution rule
in the sense of classical physics. In the opposite case of a system without
superselection sectors,
we have seen that no value attribution takes place of any observable, and we
should apparently look
in the (unsatisfactory) direction of regarding states as preparation
procedures.

In this context it is remarkable that the $\Up$ state of the pointer (taken by
itself) does not
ascribe the value $+1$ to the observable $S$, although it has this very
expectation value with
mathematical variance (spread)  0.  This should not make us feel too
uncomfortable, for one always
implicitly has to add the clause ``if it were observed'', which for us means
that it is brought into
a suitable correlation with a system (i.e., the environment) having `beables'
that are not observed,
and accordingly accommodates  classical (i.e., central) observables. We have
seen that if such a
correlation is indeed established, it is the observation of the correlated
state of the environment
(rather than the pointer state itself) that leads to conclusions about the
value of $S$. If, on the
other hand, no such correlation is established, the statement that $S$ has the
sharp expectation
value $+1$ should be  read as part of the mathematical  specification of the
state, rather than as a
value attribution. For the latter would only be defined counterfactually, and
we know from the EPR
discussion how dangerous it is to confuse counterfactual conditionals with
actual measurements in
quantum mechanics\fn{Cf.\ Healey (1989) for arguments not to assign a value to
certain observables
which are dispersion-free in a given state.}.

On the technical side, the models in the literature (see Joos and Zeh (1985),
Zurek (1993) and refs.\
therein; also cf.\ the discussion with refs.\ in   Busch et al.\ (1991))
describing how
system-environment interaction leads to decoherence, should be extended so as
to deal with very large
systems (so far, only the environment $E$ has been treated as macroscopic). In
that way, the
robustness of Zurek's stability condition can be examined. Also, stochastic
Schr\"{o}dinger equations
for the pointer state should be derived from its coupling to $E$, and solved;
much of the necessary
mathematical technology is under active development,  inspired by rather
different
philosophies, but leading to similar mathematical structures, cf.\ the
contributions of A. Amann, P.
Blanchard and A. Jadczyk, P. Bona, N. Gisin, and H. Primas\fn{{\sp  In the
approach of
Primas the entire environment is
taken to be classical, i.e., its algebra of observables is commutative. This
exorbitant truncation of
its beables is well-motivated by a deep result (Raggio's theorem) to the effect
that only in that case
the quantum system coupled to the environment is free of EPR-correlations with
it, and admits a
so-called individual description. For our purposes the mere presence of a
central subalgebra
suffices.}} to Busch et al.\ (1994), as
well as Blanchard and Jadczyk (1993) and  Jadczyk (1994a,b),
 for an up-to-date survey, and the seminal paper Gisin (1984) for older
work\fn{ {\sp  It
is to be expected that such stochastic equations, derived from the unitary
time-evolution of the
whole system,  reproduce the main features of the so-called GRW theory
(Ghirardi et al., 1986),
cf.\ Jadczyk (1994b).
Thus  it is curious that the spontaneous localization model of GRW is often
interpreted as a
fundamental modification of quantum mechanics.}}.   A theory
of human consciousness would be helpful, too.

 To close the paper, we will now recapitulate by answering some rational
objections to the
superselection programme we found in the literature. They come from van
Fraassen's
book (van Fraassen, 1991), and are attributed to Beltrametti-Cassinelli,
Hughes, Leggett, and van
Fraassen himself, respectively (in  Busch et al.\ (1991) the second one is
ascribed to Piron).
\begin{enumerate}
\item {\em Question:} what accounts for the superselection rule?\\ {\em
Answer:} Ultimately, the
locality of the observer. Under appropriate circumstances the coherence of the
coupled system is
delocalized, hence ever-present from an absolute point of view, but lost to the
observer.
This phenomenon causes most instances of `objectification' in quantum
mechanics.
 \item {\em
Question:} the fact that the system carrying the superselection rules is able
to evolve in finite
time from a given initial state into various different sectors (depending on
the state of the system
it is coupled to), implies that its Hamiltonian is not an observable. How about
that? After all, the
identification of the Hamiltonian  with the observable Energy is one of the
cornerstones of quantum
mechanics.\\
 {\em Answer:} We have seen that in Hepp's approach this objection is met by
taking the
$t\raw\infty$ limit (at least when the time-evolution is automorphic).  This
limit also played a role
in the environmental appraoch, in that the inner product $(e_n(t),e_m(t))\simeq
\dl_{nm}$ only for
$t\raw \infty$ and an infinite environment (cf.\ section 4). As we have seen,
the former limit was
fatal, but the latter harmless. But that does not answer the question, for even
if the inner product
above is nonzero for finite times, the initial environment state  must still
evolve nontrivially
`through' the sectors, and that is not possible either, if the Hamiltonian is a
function of the
observables (cf.\ the discussion in section 4). The answer to the question is
that the
evolution driving the environment through its various superselection sectors is
  not  Hamiltonian;
more importantly, this evolution is generated by operators which do not belong
to the algebra of
observables.
 This is possible and consistent because of the special way the superselection
rules of
the environment arise in our approach. Namely, we start with a simple algebra
of `beables' of the
environment; the Hamiltonian is in (or, technically, is affiliated to) this
algebra. Then we truncate
this algebra of beables to a smaller, effective algebra of observables having a
nontrivial centre.
The Hamiltonian is not in this smaller algebra, but it still drives the
time-evolution of its state
space. In other words, the reason that the Hamiltonian is not an observable is
that we have
willy-nilly restricted the set of operators to define the algebra of
observables, but the discarded
operators still contribute to the Hamiltonian.
\item {\em Question:}  Do superselection rules add
empirical content?\\ {\em Answer:} yes,  in the following sense:  the
inspection of a certain
apparatus or the study of some environment coupling to a given system may
reveal that certain
operators are not monitored, and that a `pointer basis' of the system is
singled out. In case that
this  leads to superselection rules, the theory will predict that certain
superpositions do not exist
as pure states. See Amann (1991)  for the case of chiral molecules, and  Wan
and Harrison (1993) for
Josephson junctions in SQUID rings. Thus far, such `predictions' have not quite
run ahead of
observations, but as theoretical explanations they are nontrivial. Moreover,
certain superselection
rules exist in their own right (e.g., parity), in the sense that they are not
caused by the limited
resources available to physicists and other IGUS's.  Perhaps new such rules may
be empirically
discovered.
  \item {\em Question:} Is it claimed that quantum
mechanics without superselection rules makes no predictions for what happens in
micro processes in the
iono-sphere?\\ {\em Answer:} Relative to the algebra of all beables in the
world, quantum mechanics
indeed fails to predict any concrete event or outcome.  Without superselection,
nothing `happens' up
there in the iono-sphere. But there is still the state of the total system and
its restriction to the
algebra of beables of the iono-sphere; this restriction is generically mixed,
and its time-evolution
is duly given by the theory. However, one does not need a balloon with an
observer  in it in order to
make predictions of {\em events} defined relative to any such observer; their
algebras of observables
will all be compatible with locality, and practically all events in the
iono-sphere, with their
associated probabilities,  will be meaningfully defined for all such local
algebras of observables
simultaneously. But, once again, these events are not there {\em
intrinsically}.
\end{enumerate}
We hope that the reader is satisfied with these answers. If so, we trust that
(s)he agrees with the
author, that the real problem in the interpretation of quantum mechanics is not
the explanation of
classical `reality' in a quantum world (for the interpretation of objective
classical phenomena is
rather clear), but the clarification of the physical meaning of the `kernel' of
quantum mechanics in
situations where no `objectification' in the usual sense takes place.
Thus we feel that the Copenhagen interpretation is too limited in its claim
that the entire physical
 meaning of quantum mechanics must be expressed
in terms of  the properties of classical physics\fn{Hence even the Copenhagen
interpretation has B-realist tendencies, for it attempts to relate the
subjective quantum world to a
putative objective classical world in which we can tell our friends what we
have done and what we
have learnt. A version of Copenhagen which is more in the spirit of our
proposal posits that
measurement apparatuses behave {\em as if they were classical}.}; the physical
interpretation of
non-central operators is as yet merely unknown - not meaningless in principle.

\section*{\mbox{}}
\subsection*{Acknowledgements}
This is an elaboration of a talk given at the Dept.\ of History and Philosophy
of Science,
University of Cambridge. It is a pleasure to thank J. Butterfield and M.
Redhead for the invitation
to speak, and for great moral support. I wish to thank them, as well as T.
Breuer, D. Dieks, F.
Harrison,  A. Kent, R. Omn\`{e}s, S. Saunders, M. Winnink,  and W. Zurek for
patiently explaining
their points of view to me. This paper owes a great deal to its referees, as
well as to
 T. Breuer, J. Butterfield, D. Dieks, and F.
Harrison, who commented on the first draft.
\newpage
 \section*{References}
  Aharonov, Y.,  Albert, D.Z., and  Vaidman, L. (1986)
`Measurement process in relativistic quantum theory.' {\em Phys. Rev.} {\bf
D34}, 1805-1813.\\
 Amann, A. (1991) `Chirality: a superselection rule generated by the molecular
environment?'
{\em  J.
Math. Chem.} {\bf 6},   1-15.\\
Bacciagaluppi, G. (1994) `A Kochen-Specker contradiction in the modal
interpretation of quantum
mechanics.' Univ. of Cambridge preprint.\\
Bacciagaluppi, G. and Hemmo, M. (1994) `Modal interpretations of imperfect
measurements'.
 Univ. of Cambridge preprint.\\
 Balian, R.,  Alhassid, Y., and  Reinhardt, H. (1986) `Dissipation in many-body
systems: a geometric approach based on information theory.' {\em Phys. Rep.}
{\bf 131},  1-146.\\
Barbour, J.B. (1994) `Timelessness in quantum gravity and the recognition of
dynamics in static
configurations.' {\em Class. Quantum Gravity}, to appear.\\
Bell, J.S. (1975) `On wave packet reduction in the Coleman-Hepp model.'   {\em
Helv. Phys. Acta} {\bf
48},
 93-98. \\
  Beltrametti, E.G. and  Cassinelli, G. (1981) {\em The logic of quantum
mechanics.}
 Reading: Addison-Wesley.\\
Blanchard, Ph. and Jadczyk, A. (1993) `On the interaction between classical and
quantum systems.'
{\em Phys. Lett.} {\bf A175}, 157-164.\\
  Bohm, D. (1951) {\em Quantum theory.} New York: Prentice-Hall.\\
  Bona, P. (1980)  `A solvable model of
particle detection in quantum theory.' {\em Acta F.R.N. Univ. Comen. Physica}
{\bf XX}, 65-95.\\
Bratteli,  O. and Robinson, D.W. (1987) {\em Operator Algebras and
Quantum Statistical Mechanics I}. 2nd ed. Berlin: Springer.\\
Breuer, T., Amann, A., and Landsman, N.P. (1994) `Inaccuracy and spontaneous
symmetry breaking in
quantum measurements.' {\em J. Math. Phys.} {\bf ??}, ?-?.\\
 Bub,  J. (1988)  `How to solve the measurement problem of quantum mechanics.'
{\em Found. Phys.}
{\bf  18},  701-722.\\
 Busch,  P., Lahti,  P.J., and Mittelstaedt,  P.  (1991) {\em The quantum
theory of measurement}.
 Berlin: Springer.\\
 Busch,  P., Lahti,  P.J., and Mittelstaedt,  P.,  eds.  (1994)
{\em Symposium on the Foundations of Modern Physics.}
Singapore: World Scientific.\\
Davies, E.B. ( 1976) {\em Quantum Theory of Open Systems}.
London: Academic Press. \\
  D'Espagnat, B. (1990) `Towards a separable `empirical reality?' {\em Found.
Phys.} {\bf 20},
1147-1172.\\
 Dieks, D. (1991) `On some alleged difficulties in the interpretation of
quantum mechanics.'
{\em Synthese} {\bf 86},  77-86.\\
  Dieks, D. (1994a) `Modal interpretation of quantum mechanics, measurements,
and macroscopic
behaviour.'  {\em Phys. Rev.} {\bf A49}, 2290-2300.\\
  Dieks, D. (1994b)  `Physical motivation of the modal interpretation of
quantum mechanics.' to
appear in {\em Phys. Lett.} {\bf A}.\\
   Fraassen, B.C. van (1980) {\em The Scientific Image}.  Oxford:
 Oxford University Press. \\
 Fraassen, B.C. van  (1991) {\em Quantum mechanics - an empiricist view}.
Oxford: Oxford
University Press.\\
Gell-Mann, M.  and  Hartle, J. (1990)  `Quantum mechanics in the light of
quantum
cosmology.' In: Zurek, W. (ed.), {\em Complexity, entropy, and the physics of
information}, SFI
Studies in the sciences of complexity, Vol.\ VIII (pp.\ 425-458).    Reading:
Addison-Wesley.\\
 Gell-Mann, M.  and  Hartle, J. (1993)  `Classical equations for quantum
systems.' {\em Phys. Rev.}
{\bf D47}, 3345-3382.\\
  Gisin, N. (1984)  `Quantum
measurement and stochastic processes.' {\em Phys. Rev. Lett.} {\bf 52},
1657-1660.\\
  Ghirardi, G.C.,  Rimini,  A., and Weber, T. (1986) `Unified dynamcis for
microscopic and
macroscopic systems'. {\em Phys. Rev.} {\bf D34},  470-491.\\
 Gottfried, K. (1966) {\em Quantum mechanics}.  New York: Benjamin.\\
 Haag, R. (1992) {\em Local quantum
physics}. Berlin: Springer.\\
 Haag, R. and  Kastler, D. (1964)  `An algebraic approach to
quantum field theory.'   {\em J.Math.Phys.} {\bf 5}, 848-861.\\
Healey, R. (1989) {\em The philosophy of quantum mechanics - An interactive
interpretation}.
Cambridge: Cambridge  University Press.\\
 Hepp, K. (1972)  `Quantum theory of measurement and macroscopic
observables.' {\em Helv. Phys. Acta} {\bf 45},  237-248.\\
Jadczyk, A. (1994a) `Topics in quantum dynamics.'\ BiBoS preprint 635/5/94.\\
Jadczyk, A. (1994b) `Particle tracks, events, and quantum theory.' RIMS (Kyoto)
preprint.\\
 Jauch, J.M. (1968) {\em Foundations of Quantum Mechanics}.
 Reading (Mass.): Addison-Wesley.\\
 Joos, E. and Zeh, H.D.  (1985) `The emergence of classical properties
through interaction with the environment.' {\em Z. Phys.} {\bf B59}, 223-243.\\
Kampen, N.G. van (1988) `Ten theorems about quantum mechanical
measurement.'  {\em  Physica} {\bf A153},  97-113.\\
  Khalfin, L.A. and  Tsirelson (1994), B.S. (1992)  `Quantum/classical
transition in the light of Bell's
inequalities.'  {\em Found. Phys.} {\bf 22},  879-948.\\
Kochen,  S. (1985) `A new interpretation of quantum mechanics.' In:
Lahti,  P.J. and Mittelstaedt,  P. (eds.), {\em  Symposium on the Foundations
of Modern Physics.}
(pp.\ 151-170).  Singapore: World Scientific.\\
  Landsman, N.P. (1991)  `Algebraic theory of superselection sectors and the
measurement problem
in quantum mechanics.' {\em Int. J. Mod. Phys.} {\bf A30},  5349-5371.\\
 Lloyd, S. (1988)  `Difficulty in detecting deviations from wavefunction
collapse.'
 {\em Phys. Rev.} {\bf A38},  3161-3165.\\
 Neumann, J.~von (1932) {\em Mathematische Grundlagen der
Quantenmechanik}.  Berlin: Springer-Verlag.\\
 Neumann, J.~von (1938) `On infinite direct products.' {\em Compositio Math.}
{\bf 6}, 1-77.\\
  Omn\`{e}s, R. (1992)
`Consistent interpretations of quantum mechanics.' {\em Rev. Mod. Phys.} {\bf
64},  339-382.\\
  Peres, A. (1980) `Can we undo quantum measurements?' {\em Phys. Rev.} {\bf
D22},  879-883.\\
  Peres, A. (1986) `When is a quantum measurement.' {\em  Amer. J. Phys.} {\bf
54},  688-692.\\
  Primas, H. (1990) `Induced nonlinear time evolution of open quantum objects.'
In:
 Miller,  A.I. (ed.),  {\em Sixty-two
years of uncertainty: Historical, philosophical, physics enquiries into the
foundations of quantum
phycics}. New York: Plenum.\\
 Tsirelson, B.S. (1994) `This non-axiomatizable quantum theory.' Preprint
IHES/M/94/2.\\
   Wan, K.K. (1980) `Superselection rules, quantum measurement, and
Schr\"{o}dinger's cat.'
{\em  Can.
J. Phys.} {\bf 58},  976-982.\\
Wan, K.K. and  Harrison, F. E. (1993) `Superconducting rings, superselection
rules and quantum
measurement problems.' {\em Phys. Lett.} {\bf A174}, 1-8.\\
Wan, K.K. and Jackson, T.D. (1984) `On local observables in quantum mechanics.'
 {\em Phys. Lett.} {\bf A106}, 219-223.\\
 Wheeler, J.A.  and  Zurek, W.H., eds. (1983) {\em Quantum Theory and
Measurement}.  Princeton:  Princeton University Press.\\
  Zeh, H.D. (1970) `On the interpretation of measurement in quantum theory.'
  {\em Found.Phys.} {\bf 1},  69-76.\\
 Zurek, W.H. (1981) `Pointer basis of quantum apparatus: into what mixture does
the wave
packet collapse?'   {\em  Phys. Rev.} {\bf D 24},  1516-1525.\\
  Zurek, W.H. (1982)  `Environment-induced superselection rules.'
 {\em Phys. Rev.} {\bf D26},  1862-1880.\\
  Zurek, W.H. (1993) `Preferred states, predictability, classicality and the
environment-induced decoherence.' {\em  Progr. Theor. Phys.} {\bf 89}, 281-312.
 \end{document}